\documentclass[aps,pra,preprint,groupedaddress,showpacs]{revtex4}
\usepackage{graphics}
\usepackage{bm}
\begin{document}

\title{Paraxial propagation in disclinated amorphous media}                                                                    
\author{Mohammad Mehrafarin}
\email{mehrafar@aut.ac.ir}
\author{Shima Gholam-Mirzaei} 
\affiliation{Physics Department, Amirkabir University of Technology, Tehran 15914, Iran} 
                                                               
\begin{abstract}
We study paraxial beam propagation along the wedge axis of a disclinated amorphous medium. The defect-induced inhomogeneity results in Berry phase and curvature that are affected by the induced uniaxial anisotropy. The Berry phase manifests itself as a precession of the polarization vector. The Berry curvature is responsible for the optical spin Hall effect in the disclinated medium, where beam deflection varies sinusoidally along the paraxial direction. Its application in determining the birefringence and the magnitude of the Frank vector is explained.
\end{abstract}
                                              
\maketitle
\section{Introduction} 

Amorphous optical media such as the amorphous silicon are used extensively as waveguides in optoelectronic devices \cite{Street,Dood}. Because of the strong light confinement and the sub-wavelength core size of these waveguides, light propagates paraxially along their axis. Normally defects are present in the medium, which affect its optical properties due to the elasto-optic strain. The study of paraxial propagation in defected amorphous media is, thus, important. 

Recently, we have examined paraxial propagation in screw dislocated amorphous media \cite{Mashhadi}. While dislocations are translational defects, rotational defects (disclinations) are also common in disordered media such as amorphous solids and liquid crystals. In the present work, we study paraxial beam propagation along the wedge axis of a disclinated amorphous medium.  The presence of wedge disclination in an initially homogeneous isotropic amorphous medium is shown to induce weak inhomogeneity as well as \textit{uniaxial} anisotropy due to the elsto-optic effect. The inhomogeneity causes an adiabatic variation in the direction of the wave propagation, resulting in Berry phase and curvature that are affected by the uniaxial anisotropy. The geometric Berry phase \cite{Berry} acquired by an optical beam (as the Pancharatnam phase \cite{Pancharatnam} or a spin redirection phase \cite{Tomita,Chiao}) has attracted extensive attention. In particular, observable effects such as the Rytov-Vladimirskii rotation \cite{Rytov,Vladimirskii} and the optical spin Hall (or Magnus) effect \cite{Zeldovich,Zeldovich2} have been derived as manifestations of Berry phase and curvature, respectively \cite{ Bliokh,Bliokh1,Bliokh2,Onoda,Onoda2,Sawada,Mehrafarin}. Here, the Berry phase manifests itself as a precession of the polarization vector, which is characteristic of anisotropic media \cite{Bliokh5}. The Berry curvature enters the equations of motion of the beam and is responsible for the opposite deflections of the right/left circularly polarized beams.  Because of the anisotropy, these deflections vary sinusoidally along the paraxial direction. This yields the optical spin Hall effect in the disclinated medium, whose application in determining the birefringence and the magnitude of the Frank vector is explained.

\section{The effect of wedge disclination}

We consider a monochromatic circularly polarized wave propagating paraxially in a homogeneous isotropic medium of refractive index $n$. The unit wave vector $\hat{\bm{k}}$ thus holds an angle $\theta$, which is always sufficiently small, with the paraxial direction $z$.  Denoting the polarization vectors by $\bm{\epsilon}_\sigma$, where
$\sigma=\pm1$  correspond to right/left circular polarization ($\bm{\epsilon}_\sigma^\dagger\bm{\epsilon}_{\sigma'}=\delta_{\sigma\sigma'}$), we have
$$
\bm{\epsilon}_\sigma=\frac{1}{\surd{2}}(\hat{\bm{\theta}}-i\sigma \hat{\bm{\varphi}})
$$
$\hat{\bm{\theta}}, \hat{\bm{\varphi}}$ being the spherical unit vectors orthogonal to $\hat{\bm{k}}$.
The beam's spin angular momentum along the direction of propagation (the helicity) is \cite{Berry3}
\begin{equation}
\bm{\epsilon}_\sigma^\dagger (-i \hat{\bm{k}}\times\bm{\epsilon}_\sigma)=\sigma. \label{hel}
\end{equation}

We consider the effect of introducing a wedge disclination, with Frank vector $\bm{\omega}$ oriented along the paraxial direction, in the initially homogeneous isotropic medium. From the standpoint of a Volterra process, the wedge disclination corresponds to cutting or inserting a material wedge of dihedral angle $\omega=|\bm{\omega}|$, which is generally small. The corresponding displacement vector field, $\bm{u}$, in cylindrical coordinates has the nonzero component $u_\varphi=(\alpha-1)\rho\varphi$, where $\alpha-1=\pm \omega/2\pi$ and the $+$($-$) sign pertains to insertion (cut). Since $\nabla\cdot\bm{u}=\alpha-1$ is small, the disclination produces slight expansion/compression and renders the medium weakly inhomogeneous. This will cause an adiabatic variation in the direction of the wave  propagation resulting in Berry phase and curvature. Furthermore, the disclination strain tensor field has the following nonzero components: 
$$S_{\rho\varphi}=S_{\varphi\rho}=\frac{1}{2}(\alpha-1)\varphi,\ \ S_{\varphi\varphi}=\alpha-1.$$
The relative permittivity tensor $n^2 \delta_{ij}$, thus, acquires an anisotropic part $\Delta_{ij}$ due to the strain, where (see e.g. \cite{Liu}) 
$$
\Delta_{ij}=-n^4 p_{ijkl}S_{kl}
$$ 
$p_{ijkl}$ being the elasto-optic coefficients of the medium. A wedge disclination, therefore, renders an otherwise homogeneous isotropic medium weakly inhomogeneous and anisotropic. For amorphous media, where only two independent elasto-optic coefficients (customarily denoted by $p_{11}$ and $p_{12}$) exist, we find, after detailed calculations,
$$
\Delta_{\rho\rho}=\Delta_{zz}=-(\alpha-1)p_{12}n^4, \ \ \Delta_{\varphi\varphi}=-(\alpha-1)p_{11}n^4
$$
other components being zero. The principle refractive indices are, therefore,
$$
n_\rho= n_z=n-\frac{1}{2}(\alpha-1)p_{12}n^3,\ \ n_\varphi=n-\frac{1}{2}(\alpha-1)p_{11}n^3
$$
to first order in the elasto-optic perturbation. (The adiabatic variation of refractive indices with position are to be ignored, of course.) The weak \textit{uniaxial} anisotropy thus induced in the amorphous medium results in a phase difference for the two linearly polarized modes that constitute the paraxial beam. Therefore, the polarization vector becomes
\begin{equation}
\bm{\epsilon}_\sigma =\frac{1}{\surd 2}(\hat{\bm{\theta}}- i \sigma e^{ik_0\Delta n z} \hat{\bm{\varphi}}) \label{e}
\end{equation}
where $k_0$ is the wave number in vacuum and
\begin{equation}
\Delta n=n_\varphi-n_\rho=\frac{1}{2}(\alpha-1) (p_{11}-p_{12})n^3 \label{biref}
\end{equation}
is the induced birefringence. Note that $\bm{\epsilon}_\sigma$ still satisfy the orthonormality condition $\bm{\epsilon}_\sigma^\dagger\bm{\epsilon}_{\sigma'}=\delta_{\sigma\sigma'}$, of course. The beam's helicity is calculated from (\ref{hel}) to be $\sigma \cos(k_0\Delta nz)$. As expected \cite{Berry3}, the helicity varies along the paraxial direction due to the induced elasto-optic birefringence and reduces to the constant value $\sigma$ in the absence of the wedge disclination ($\alpha=1$). 

\section{Berry effects in the beam dynamics}

The adiabatic variation of the refractive indices with position has negligible dynamical effect and was, therefore, ignored. However, the resulting adiabatic variation of the beam direction $\hat{\bm{k}}$ plays a geometric role with nontrivial consequences for the beam dynamics. As usual, the variation gives rise to a parallel transport law in the momentum space, defined by the Berry connection (gauge potential)
$$
\bm{A}_{\sigma \sigma'}(\bm{k})=\bm{\epsilon}_\sigma^\dagger(-i\nabla_{\bm{k}}) \bm{\epsilon}_{\sigma'}. 
$$
Using (\ref{e}), we obtain
$$
\bm{A}_{\sigma \sigma'}=\left(\cos(k_0\Delta nz)\delta_{\sigma \sigma'}+i \sin(k_0\Delta nz)(\delta_{\sigma \sigma'}-1)\right)\sigma\frac{\cot\theta}{k} \hat{\bm{\varphi}}
$$
or in matrix notation,
\begin{equation}
\bm{A}= (\bm{\sigma} \cdot\bm{h})\frac{\cot\theta}{k} \hat{\bm{\varphi}} \label{m}
\end{equation}
where $\bm{\sigma}$ is the Pauli matrix vector and
$$
\bm{h}(z)=(0,\sin(k_0\Delta nz),\cos(k_0\Delta nz)).
$$
Equation (\ref{m}) describes the parallel transport of the polarization vector along the beam and generalizes a previous result for inhomogeneous isotropic media \cite{Bliokh,Bliokh1,Bliokh2}. The Berry curvature (gauge field strength) associated with this connection is ($\bm{A} \times \bm{A}=0$)
$$
\bm{B}=\nabla_{\bm{k}} \times \bm{A}=-(\bm{\sigma} \cdot \bm{h})\frac{\bm{k}}{k^3}. 
$$

In the course of propagation, the polarization evolves according to $\bm{\epsilon}_\sigma \rightarrow e^{i\Theta}\bm{\epsilon}_\sigma$, where  
\begin{equation}
\Theta=\int_C \bm{A} \cdot d\bm{k}= (\bm{\sigma} \cdot \bm{h})\Theta_0 \label{ph}
\end{equation}
is the geometric Berry phase. Here $C$ is the beam trajectory in momentum space and $\Theta_0=\int_C \cos \theta d\varphi$ is the Berry phase accumulated for $\sigma=1$ in the absence of anisotropy. (In the absence of anisotropy, (\ref{ph}) simply yields the phase factor $e^{i\sigma\Theta_0}$ that leads to the well known Rytov rotation.) The evolution, thus, entails a  precession of the polarization vector, which is characteristic of anisotropic media \cite{Bliokh5}, about the unit vector $\bm{h}$. In view of the polarization evolution, the Berry curvature for a given beam is, therefore,
$$
\bm{B}_\sigma=(e^{i\Theta}\bm{\epsilon}_\sigma)^\dagger\bm{B} (e^{i\Theta}\bm{\epsilon}_\sigma)=
\bm{\epsilon}_\sigma^\dagger\bm{B}\bm{\epsilon}_\sigma
$$
where the last expression follows because $\Theta$ and $\bm{B}$ commute. Hence
$$
\bm{B}_\sigma=-\sigma \cos(k_0\Delta nz)\frac{\bm{k}}{k^3}
$$
which reduces to the well known result in the absence of anisotropy, namely, the field of a magnetic monopole of charge $\sigma$ situated at the origin of the momentum space \cite{Bliokh,Bliokh1,Bliokh2}. 

\begin{figure}
\includegraphics{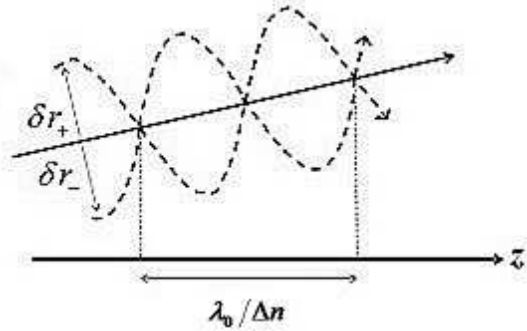}
\caption{
Propagation of oppositely polarized paraxial beams in a wedge disclinated amorphous medium. In the absence of disclination, the deflections vanish and the two trajectories collapse along the solid line.}
\label{fig1}
\end{figure}

The equations of motion of the beam in the presence of momentum space Berry curvature have been derived repeatedly for various particle beams (photons \cite{Bliokh,Bliokh1,Bliokh2,Zeldovich2,Onoda,Onoda2,Sawada,Mehrafarin}, phonons \cite{Bliokh6,Torabi,Mehrafarin2} and electrons \cite{Chang,Sundaram,Culcer,Berard}). The beam trajectory, $\bm{r}_\sigma$, satisfies
$$
\dot{\bm{r}}_\sigma=\hat{\bm{k}}+\bm{B}_\sigma\times \dot{\bm{k}}\nonumber
$$
where dot denotes derivative with respect to the beam length. This differs from the standard ray equation of the geometrical optics, which holds in the absence of disclination, by the term involving the Berry curvature. It yields the beam deflection
$$
\delta \bm{r}_\sigma(z)=-\sigma \cos(k_0\Delta nz)\int_C \frac{\bm{k}}{k^3}\times d\bm{k} 
$$
which results in the splitting of beams of opposite polarizations and, being orthogonal to the beam direction, produces a spin current across the direction of propagation. This is the optical spin Hall effect in the disclinated medium and generalizes a previous result for inhomogeneous isotropic media \cite{Bliokh,Bliokh1,Bliokh2}. $\delta\bm{r}_\sigma$ varies sinusoidally along the paraxial direction with wavelength $\lambda_0/\Delta n$, where $\lambda_0$ is the beam's wavelength in vacuum (see figure \ref{fig1}). In particular, it vanishes for successive beam points that are separated by $\lambda_0/2\Delta n$ along the $z$-axis. Measurement of this determines the birefringence and provides an indirect method for determining the magnitude of the Frank vector, $\omega$, through (\ref{biref}).

\end{document}